\documentclass[a4paper,11pt]{article}
\pdfoutput=1 % if your are submitting a pdflatex (i.e. if you have
\usepackage{jcappub} % for details on the use of the package, please
\title{Noncanonical warm inflation with nonminimal derivative coupling}

\author[a,1]{Xiao-Min Zhang,\note{Corresponding author.}}
\author[a]{Run-Qing Zhao,}
\author[b,2]{Zhi-Peng Peng,\note{Corresponding author.}}
\author[c,3]{Xi-Bin Li,\note{Corresponding author.}}
\author[a]{Yun-Cai Feng,}
\author[a]{Peng-Cheng Chu}
\author[a]{and Yi-Hang Xing}
\affiliation[a]{School of Science, Qingdao University of Technology, \\ Qingdao, 266520 China}
\affiliation[b]{School of Physics, Henan University of Technology, \\ Zhengzhou 450001, China}
\affiliation[c]{College of Physics and Electronic Information, Inner Mongolia Normal University, \\ Hohhot, 010022, China}
\emailAdd{zhangxm@mail.bnu.edu.cn}
\emailAdd{18733666528@163.com}
\emailAdd{zhipeng@mail.bnu.edu.cn}
\emailAdd{lxbbnu@mail.bnu.edu.cn}
\emailAdd{fengyuncai@qut.edu.cn}
\emailAdd{kyois@126.com}
\emailAdd{3415769493@qq.com}
\abstract{This study extended noncanonical warm inflation to the nonminimal derivative coupling scenario. The fundamental equations, including the evolution equations and the slow roll equations of this new framework, were derived. The enlarged damping term, which encompasses both gravitationally enhanced friction and thermal damping, resulted in a well overdamped inflationary process, ensuring that the slow roll approximations can be satisfactorily satisfied. A linear stability analysis corroborated the viability of this approach, yielding significantly relaxed slow roll conditions within the context of noncanonical warm inflation with nonminimal derivative coupling. Subsequently, the density fluctuations in this new framework were analyzed, leading to approximately analytic results for the power spectrum, spectral index, and related quantities. Both the energy scale at horizon crossing and the tensor-to-scalar ratio decreased considerably because of the effects of thermal damping and nonminimal derivative coupling. The upper bound for field excursion remained safely sub-Planckian in this inflationary scenario. Thus we reached a successful and meaningful model to broad the scope of warm inflation.}

\keywords{warm inflation, nonminimal derivative coupling, stability analysis, cosmological perturbations}
\arxivnumber{2410.16839}

\begin{document}
\maketitle
\section{\label{sec:level1}Introduction}
Inflation, characterized by a quasi-exponential accelerated expansion in the very early universe, effectively addresses key cosmological issues such as the horizon, flatness, and monopole problems \cite{Guth1981,Linde1982,Albrecht1982,Bassett2006}. It also provides a mechanism for generating the seeds of large-scale structures and the observed small anisotropies in the cosmic microwave background (CMB) through vacuum fluctuations \cite{PLANCK1,PLANCK2}. Inflationary theory is generally divided into two types: standard (or cold) inflation and warm inflation. Warm inflation, first introduced by Berera and Fang \cite{BereraFang,Berera1995}, suggests that radiation is continuously produced through interactions $\mathcal{L}_{int}$ between the inflaton field and other sub-dominant bosonic or fermionic fields, thereby eliminating the need for a distinct reheating phase. This enables a smooth transition to the radiation-dominated phase of the universe. In warm inflation, density fluctuations arise primarily from thermal rather than vacuum fluctuations \cite{BereraFang,Lisa2004,Berera2000}, a notable distinction from cold inflation. The thermal damping effect relaxes the flatness requirements of the potential, making the slow-roll conditions easier to achieve \cite{Ian2008,Campo2010,ZhangZhu}. Consequently, issues such as the $\eta$ problem \cite{etaproblem,BereraIanRamos} and the large inflaton amplitudes \cite{Berera2005,BereraIanRamos}, which are prominent in cold inflation, are mitigated.

Inflation is typically modeled using a canonical scalar field with a Lagrangian density of the form $\mathcal{L}=X-V$, where $X=\frac12 g^{\mu\nu}\partial_{\mu}\phi\partial_{\nu}\phi$, and $V$ denotes the inflaton potential. To explore more diverse inflationary scenarios, studies have proposed various models, including those involving noncanonical fields, pure kinetic fields, multi-field models \cite{refining2012,Mukhanov2006,Armendariz-Picon,Garriga1999,Gwyn2013,Tzirakis,Franche2010,Eassona2013,Bean2008,multifield}, and nonminimal coupling fields \cite{Kaiser1995,Karydas2102.08450,GermaniPRL2010}, In the past three decades, warm inflation has also undergone significant developments, particularly in microphysical research \cite{Berera2016,MossXiong}, cosmological perturbations \cite{Berera2000,Zhang2014,WangYY2019,Berera2016,Zhang2023,WarmSPy2024} and model extensions \cite{Zhang2014,Zhang2018,Peng2016,WangYY2018,EadkhongNPB2023}. While most warm inflation models involve canonical fields, exceptions such as the warm Dirac-Born-Infeld (DBI) inflationary model \cite{Cai2011}, exist. Our previous work has proposed general noncanonical warm inflation models \cite{Zhang2014,Zhang2018,Zhang2021}, warm $k$-inflation \cite{Peng2016,Peng2018,Zhang2023}, tachyon warm inflation \cite{ZhangTachyon} and two-field warm inflation models \cite{WangYY2018,WangYY2019}. Noncanonical fields exhibit several novel features, including second-order equations of motion and relaxed slow-roll conditions compared to canonical inflationary theory \cite{refining2012}. Additionally, the energy scale of inflation is lower at the time of horizon crossing, allowing it to be safely described by effective field theory. Furthermore, the tensor-to-scalar ratio decreases significantly in most noncanonical models \cite{refining2012,Cai2011,Zhang2014} although some phenomenological models exhibit an increase in this ratio \cite{Mukhanov2006}.

Most warm inflationary models have been constructed within the framework of general relativity (GR). This study focuses on extending warm inflation to the framework of nonminimal coupling theory, broadening its scope. In nonminimal coupling theory, two primary scenarios are considered: the general $f(\phi)R$ case, wherein the inflaton field couples directly to the Ricci scalar \cite{Kaiser1995,Karydas2102.08450,EadkhongNPB2023}, and the nonminimal derivative coupling (NMDC) case, wherein the inflaton kinetic term is coupled nonminimally to the gravitational term \cite{Karydas2102.08450,GermaniPRL2010,DalianisJCAP2020,Sadjadi2015}. Although NMDC inflation models have primarily been studied within the context of cold inflation, this paper extends noncanonical warm inflation proposed in \cite{GermaniPRL2010}, to the NMDC framework. In this scenario, the derivative term of the inflaton field is nonminimally coupled to the Einstein tensor. More general cases of nonminimal derivative coupling models for scalar fields are discussed in \cite{Asimakis2023,Karydas2102.08450,DalianisJCAP2020}. In this study, the nonminimal derivative coupling term is specified as $G^{\mu\nu}\partial_{\mu}\phi\partial_{\nu}\phi$ \cite{GermaniPRL2010}, ensuring that the field equations involve no more than second derivatives and that gravitationally enhanced friction slows the evolution of the scalar field. When the kinetic term of the scalar field is nonminimally coupled to the Einstein tensor, the effective self-coupling $\lambda$ of the Higgs boson can be reduced to order $1$, and this new Higgs inflation model introduces no additional degrees of freedom \cite{GermaniPRL2010,YangNan2015}. For a massless scalar field without a canonical kinetic term $g^{\mu\nu}\partial_{\mu}\phi\partial_{\nu}\phi$, the nonminimally derivative-coupled scalar field exhibits behavior analogous to dark matter \cite{Gao2010,Ghalee2013}. The introduction of nonminimal coupling gravitational friction or thermal effects which generate a friction term $\Gamma\dot\phi$ allows the $\lambda\phi^4$ potential case previously excluded in minimal coupling standard inflation to align with observational data \cite{YangNan2015,BeingWarm2014}. Previous works \cite{Zhang2014,Zhang2018} have proposed noncanonical warm inflation models. In this paper, we conduct a systematic analysis of noncanonical warm inflationary models incorporating nonminimal derivative coupling. Our findings indicate that the predictions of the new model align well with observations and provide several distinct advantages. The most notable feature is the substantial friction in the evolution equation, which includes the enhanced Hubble damping term, the NMDC gravitational friction term, and the thermal damping term. This combination allows the slow-roll approximations to remain valid easily independent of the specific potentials involved. Additionally, the Lyth bound in general relativity is modified to a smaller value, ensuring that the field excursion remains well below the Planck scale.

The structure of this paper is as follows: Section \ref{sec:level2} presents the construction of the new NMDC noncanonical warm inflationary scenario and derives the fundamental equations governing the model. Section \ref{sec:level3} introduces the slow-roll approximations for this model and includes a comprehensive linear stability analysis to establish the conditions under which the slow-roll approximation holds. Section \ref{sec:level4} is dedicated to a systematic analysis of cosmological perturbations within the model. Finally, Section \ref{sec:level5} offers discussions and conclusions.

\section{\label{sec:level2}NMDC noncanonical warm inflationary scenario}

Based on the framework in \cite{GermaniPRL2010}, the unique nonminimally derivative-coupled Higgs theory to gravity, which propagates no additional degrees of freedom beyond those of general relativity minimally coupled to a scalar field, can be described by the following equation. Considering all components of the universe, the total action in the multi-component NMDC warm inflationary scenario is expressed as follows:
\begin{equation}\label{action}
    S=\int d^4x\sqrt{-g}\left[\frac{1}{2}M_p^2R+\mathcal{L}(X,\phi)+\frac{G^{\mu\nu}}{2M^2}\partial_
    {\mu}\phi\partial_{\nu}\phi+\mathcal{L}_{int}+\mathcal{L}_R\right],
\end{equation}
where $G_{\mu\nu}$ represents the Einstein tensor, $M_p^2 \equiv {\left( {8\pi G} \right)^{ - 1}}$ is the reduced squared Planck mass, and $M$ is the coupling constant with the dimension of mass. The total Lagrangian density is expressed as $\mathcal{L}_{total}=\mathcal{L}_g+\mathcal{L}(X,\phi)+\mathcal{L}_{NMDC}+\mathcal{L}_R+\mathcal{L}_{int}$, where $\mathcal{L}_g=\frac{1}{2}M_p^2R$ denotes the Lagrangian density for the gravitation, and $\mathcal{L}_{NMDC}=\frac1{2M^2}G^{\mu\nu} \partial_{\mu}\phi\partial_{\nu}\phi$ represents the NMDC Lagrangian density. The Lagrangian density for the noncanonical inflaton field is $\mathcal{L}_{noncan}= \mathcal{L}(X,\phi)$, a general function of the field $\phi$ and the kinetic term $X$, and for simplicity, $\mathcal{L}$ is used to denote $\mathcal{L}(X,\phi)$ hereinafter. $\mathcal{L}_R$ describes the radiation fields, while $\mathcal{L}_{int}$ accounts for the interactions between the inflaton and all other subdominant components. The Lagrangian density must reduce to the canonical case (i.e. $\mathcal{L}= X-V$) in the small $X$ limit to maintain a uniform normalization of the field.

A proper noncanonical Lagrangian density must satisfy the conditions: $\mathcal{L}_{X}\geq0$ and $\mathcal{L}_{XX}\geq0$ (where the subscript $X$ denotes a derivative, while the subscripts in $\mathcal{L}_g$, $\mathcal{L}_R$ and $\mathcal{L}_{int}$ serve only as labels). These conditions ensure compliance with the null energy condition and allow for the physical propagation of perturbations \cite{Franche2010,Bean2008}. From these conditions and the field normalization, the expression for $\mathcal{L}_X\geq1$ can be obtained. The evolution equation for the inflaton field can be obtained by varying the action as follows:
\begin{equation}\label{EOM0}
\frac{\partial(\mathcal{L}(X,\phi)+\mathcal{L}_{int})}{\partial\phi}-\left(\frac{1}{\sqrt{-g}}
\right)\partial_{\mu}\left(\sqrt{-g}\frac{\partial\left(\mathcal{L}(X,\phi)+\mathcal{L}_{NMDC}\right)}
{\partial(\partial_{\mu}\phi)}\right)=0.
\end{equation}

In the spatially flat Friedmann-Robertson-Walker (FRW) universe, the inflaton mean field is assumed to be homogeneous, i.e. $\phi=\phi(t)$. Hence, the motion equation reduces to

\begin{eqnarray}\label{EOM1}
    \left[\left(\frac{\partial\mathcal{L}(X,\phi)}{\partial X}\right)+2X\left(\frac{\partial^2\mathcal{L}(X,\phi)}{\partial X^2}\right)+\frac{3H^2}{M^2}\right]\ddot{\phi}  +\frac{\partial^2\mathcal{L}(X,\phi)}{\partial X\partial\phi}\dot\phi^2\nonumber\\+\left[3H\left(\frac{\partial\mathcal{L}(X,\phi)}{\partial X}+\frac{3H^2}{M^2}+\frac{2\dot H}{M^2}\right)\right]\dot{\phi} -\frac{\partial(\mathcal{L}(X,\phi)+
    \mathcal{L}_{int})}{\partial\phi}=0,
\end{eqnarray}
where $X=\frac12\dot\phi^2$ in the FRW universe.
By varying the action with respect to the metric, the field equations are obtained as follows:
\begin{equation}
    G_{\mu\nu}=\frac{1}{M_P^2}\left[T_{\mu\nu}^{(0)}-\frac1{M^2}T_{\mu\nu}^{(1)}\right],
\end{equation}
\begin{equation}
    T_{\mu\nu}^{(0)}=\nabla_{\mu}\phi\nabla_{\nu}\phi-g_{\mu\nu}\left(\frac{1}{2}(\nabla\phi)^2+V(\phi)\right) ,
\end{equation}
and
\begin{equation}
    \begin{aligned}T_{\mu\nu}^{(1)}&=-G_{\mu\nu}\nabla_{\lambda}\phi\nabla^{\lambda}\phi+4R^{\lambda}{}_{(\mu}\nabla_{\nu)}
    \phi\nabla_{\lambda}\phi-\nabla_{\mu}\phi\nabla_{\nu}\phi R\\&+2[\nabla^{\kappa}\phi\nabla^{\lambda}\phi R_{\mu\kappa\nu\lambda}+\nabla_{\mu}\nabla^{\lambda}\phi\nabla_{\nu}\nabla_{\lambda}\phi-\nabla_{\nu}\nabla_{\mu}\phi
    \nabla^{2}\phi]\\&+g_{\mu\nu}[\nabla^{2}\phi\nabla^{2}\phi-\nabla_{\kappa}\nabla_{\lambda}\phi\nabla^{\kappa}
    \nabla^{\lambda}\phi-2R_{\kappa\lambda}\nabla^{\kappa}\phi\nabla^{\lambda}\phi]. \end{aligned}
\end{equation}
$T_{\mu\nu}^{(1)}$ represents the additional correction to the energy-momentum tensor due to the nonminimal derivative coupling.

An important parameter for the noncanonical field is the sound speed, which characterizes the propagation speed of scalar perturbations: $c_{s}^{2}=p_{X}(\phi,X)/\rho_{X}(\phi,X)=\left(1+2X\mathcal{L}_{XX}/\mathcal{L}_{X}\right)^{-1}$, where the subscript $X$ denotes a derivative.

Now we consider the general case where the Lagrangian density has a separable form, consisting of a kinetic term and a potential term, denoted as $ \mathcal{L}=K(X)-V(\phi)$, where $K$ is the noncanonical kinetic term that depends weakly or not at all on $\phi$ \cite{Franche2010}. In this scenario, $K$ is assumed to be solely a function of $X$, leading to $\mathcal{L}_{X\phi}=0$ and $K_{X}=\mathcal{L}_{X}$. The general form of the Lagrangian can be expressed in two types: a series-form or a closed-form \cite{Franche2010}. The latter can reduce to canonical or DBI
inflation in a specific gauge $\mathcal{L}_X=c_s^{-1}$ \cite{Bean2008,Zhang2015}. The interaction term $\mathcal{L}_{int}$ in Eqs. (\ref{action}) and (\ref{EOM0}) depends on the zeroth order of the inflaton and other fields, but not on their derivatives. Effective models for the interaction between the inflaton and other fields include the supersymmetric two-stage mechanism \cite{BereraKephart,MossXiong} and the warm little inflaton mechanism \cite{Berera2016}. The term $\Gamma\dot\phi$ generally describes the dissipation of $\phi$ into other fields \cite{BereraFang,Berera2005,Berera2000,BereraIanRamos}, serving as a thermal damping term. The other terms that do not contain $\dot\phi$ in $\partial\mathcal{L}_{int}/\partial\phi$ of Eq. (\ref{EOM1}) and $\partial\mathcal{L}(X,\phi)/\partial\phi$ are resummed into the effective potential $V_{eff}$, which includes thermal corrections and depends on both the inflaton and the temperature. The temperature corresponds to that of the radiation bath, which remains nonzero due to the inflaton's dissipation into the bath, provided that the temperature dependence of the dissipation coefficient satisfies the condition derived in Eq. (\ref{c}) in Section \ref{sec:level3}. Under these assumptions, the equation of motion can be rewritten as follows:
\begin{equation}\label{EOM2}
    (\mathcal{L}_{X}c_{s}^{-2}+3F)\ddot{\phi}+3H\left(\mathcal{L}_{X}+3F+\frac{2\dot{F}}{H}\right)\dot{\phi}
    +\Gamma\dot{\phi}+V_{eff,\phi}=0,
\end{equation}
where $F=\frac{H^{2}}{M^{2}}$, and the term $\dot F\ll HF$ during inflationary epoch. Hence, the term $\frac{2\dot{F}}{H}\dot{\phi}$ in the above equation can be neglected in comparison to the preceding term. The Einstein GR limit is recovered when $F\ll1$, and $F\gg1$ indicates a regime of high gravitational friction limit.
For simplicity, the thermal effective potential $V_{eff}$ is denoted as $V$ hereinafter, and the subscript $\phi$ denotes a derivative. The Hubble damped term is $\mathcal{L}_X$ times larger than that in canonical inflation. In this case, gravitationally enhanced friction and thermal friction are introduced, slowing the evolution of the inflaton field.

By combining the Hamiltonian constraint with the radiation contribution, the Friedmann equation for the NMDC inflationary scenario is expressed as follows:
\begin{equation}
    3H^{2}=\frac{1}{M_{p}^{2}}(2X \mathcal{L}_{X}-\mathcal{L}+9FX+Ts).
\end{equation}

Considering the inflaton evolution equation, total energy conservation equation $\dot\rho+3H(\rho+p)=0$ and the total energy density $\rho=2X \mathcal{L}_{X}-K(X)+V(\phi,T)+9FX+Ts$, the entropy production equation is obtained as follows:
\begin{equation}\label{entropy1}
    T\dot{s}+3HTs=\Gamma\dot{\phi}^{2}.
\end{equation}
If thermal corrections to the potential are sufficiently small, which is shown to hold in the slow-roll regime discussed in Section \ref{sec:level3}, the radiation energy can be expressed as $\rho_r=3Ts/4$. Therefore, Eq. (\ref{entropy1}) becomes equivalent to the equation governing the production of radiation energy density $\dot{\rho}_r+4H\rho _r=\Gamma \dot{\phi}^2$.

Inflation theory must be predictive; thus, by neglecting the highest-order terms in Eqs. (\ref{EOM2}) and (\ref{entropy1}), the slow-roll approximation equations are derived:
\begin{equation}\label{EOM3}
    3H(\mathcal{L}_X+3F)\dot{\phi}+\Gamma\dot{\phi}+V_\phi=0,
\end{equation}
\begin{equation}\label{entropy2}
    3HTs=\Gamma\dot{\phi}^{2}.
\end{equation}
The damping term is enhanced by a factor of ($\mathcal{L}_{X}+3F+r$) compared to minimal coupling in standard canonical inflation, indicating that NMDC noncanonical warm inflation is strongly overdamped. The quantity $r=\Gamma/3H$ is a traditional parameter describing the thermal dissipative strength in warm inflation.

The number of e-folds in NMDC noncanonical warm inflation is given by
\begin{equation}
N=\int H dt=\int\frac{H}{\dot{\phi}}d\phi\simeq-\frac{1}{M_p^2}\int_{\phi_{\ast}}
    ^{\phi_{end}}\frac{V(\mathcal{L}_X+3F+r)}{V_{\phi}}d\phi,
\end{equation}
where $\phi_{\ast}$ refers to the field value at the Hubble horizon crossing. Due to the overdamped feature and easily satisfied slow-roll conditions of this inflationary model, the number of e-folds is sufficient to address the three major problems of the standard inflationary model.

\section{\label{sec:level3}Stability analysis}

To conduct a systematic stability analysis, we first introduce slow-roll parameters relevant to warm inflation:
\begin{equation}
 \epsilon =\frac{M_p^2}{2}\left(\frac{V_{\phi}}{V}\right) ^2,~~\eta =M_p^2\frac {V_{\phi \phi}}{V},~~ \beta
=M_p^2\frac{V_{\phi}\Gamma_{\phi}}{V\Gamma},
\end{equation}
and two additional parameters related to temperature dependence:
\begin{equation}
 b=\frac {TV_{\phi T}}{V_{\phi}},~~~ c=\frac{T\Gamma_T}{\Gamma}.
\end{equation}

We define $u=\dot{\phi}$, and the strict Eqs. (\ref{EOM2}) and (\ref{entropy1}) are rewritten as follows:
\begin{equation}\label{dotu}
\dot{u}=-(\mathcal{L}_{X}c_s^{-2}+3F)^{-1}[3H(\mathcal{L}_{X}+3F)u+\Gamma u+V_{\phi}],
\end{equation}
\begin{equation}\label{dots}
\dot{s}=-3Hs+\frac{\Gamma u^{2}}{T}.
\end{equation}

Inflation is typically associated with slow-roll approximations, which involve neglecting the highest-order terms in Eqs. (\ref{dotu}) and (\ref{dots}). The slow-roll conditions imply that the inflaton's energy is dominated by the potential, its evolution is gradual, and radiation production is quasi-static.

The variables $u_0$ and $s_0$ represent slow-roll solutions that satisfy the following equations:
\begin{equation}\label{SREOM}
    [3H(\mathcal{L}_X+3F)+\Gamma]u_0+V_{\phi}(\phi,T)=0,
\end{equation}
\begin{equation}\label{SRentropy}
    3H_0T_0s_0-\Gamma u_0^2=0.
\end{equation}

Varying the Friedmann equation yields
\begin{equation}
\delta H=\frac{(\mathcal{L}_{X}c_{s}^{-2}+9F)u\delta u+V_{\phi}\delta \phi+T\delta s}{6H(M_{p}^{2}-3\frac{X}{M^{2}})}.
\end{equation}
Considering variations in key quantities, we derive useful equations for conducting the stability analysis:
\begin{equation}
\frac{\delta\Gamma}{\Gamma_{0}}=\left(\frac{cV_{\phi}}{3T_{0}s_{0}}b+\frac{1}{M_{p}^{2}}\frac{V_{0}}{V_{\phi}}\beta\right)
    \delta\phi+\frac{c}{3s_{0}}\delta s,
\end{equation}
\begin{equation}
\frac{\delta V_{\phi}}{V_{\phi}}=\left(\frac{V_{\phi}}{3T_{0}s_{0}}b^{2}+\frac{1}{M_{p}^{2}}\frac{V_{0}}{V_{\phi}}\eta\right)\delta
    \phi+\frac{b}{3s_{0}}\delta s.
\end{equation}
Based on the thermal relation $s\simeq-V_T$, we have $\delta s=-V_{TT}\delta T-V_{\phi T}\delta \phi$ and $\delta T=\frac{1}{3s}(T\delta s+V_{\phi} b\delta\phi)$.

The exact solutions $u$, $\phi$ and $s$ are expanded around the slow-roll solutions: $u=u_0+\delta u,$
$\phi=\phi_0+\delta\phi$, $s=s_0+\delta s$. The perturbation terms $\delta u,$ $\delta s$, and $\delta \phi$ are considerably smaller than the background terms $u_0$, $s_0$ and $\phi_0$. Our linear stability analysis focuses on these slow-roll solutions to determine the conditions under which they act as formal attractor solutions for the dynamical system.

Using the new variable $u$, we have $X=\frac12u^2$, then $\delta X=u\delta u$, $\delta \mathcal{L}_X=\mathcal{L}_{XX}u\delta u$, and $\delta F=2\frac{F}{H}\delta H$. By applying the definitions of the slow-roll parameters and varying Eqs. (\ref{dotu}) and (\ref{dots}), we obtain:
\begin{equation}  \left(
\begin{array}{c} \delta \dot \phi \\ \delta \dot u \\  \delta \dot s
\end{array} \right)  =E\cdot \left(\begin{array}{c} \delta \phi \\
 \delta u \\  \delta s \end{array} \right) -F.
\end{equation}
The matrices $E$ and $F$ can be expressed as follows:
\begin{equation}
E=\left(\begin{array}{ccc} 0&1&0\\ A&\lambda_1 &B\\ C&D&\lambda_2
\end{array} \right),~~~F=\left(\begin{array}{c} 0\\ \dot u_0 \\ \dot s_0 \end{array}
\right).
\end{equation}
The elements of matrix $E$ are derived through calculations:
\begin{eqnarray}\label{A}
A=3H^{2}\left[\frac{\mathcal{L}_X+9F}{(\mathcal{L}_Xc_s^{-2}+3F)(1-\frac{3X}{M_{p}^{2}M^{2}})(\mathcal{L}_{X}+3F+r)}\epsilon
-\frac{\eta}{\mathcal{L}_{X}c_s^{-2}+3F}\right.\nonumber\\ \left.+\frac{\mathcal{L}_{X}+3F+r}{\mathcal{L}_{X}c_s^{-2}+3F}bc+\frac{r}
{(\mathcal{L}_{X}c_s^{-2}+3F)(\mathcal{L}_{X}+3F+r)}\beta -\frac{3(\mathcal{L}_{X}+3F+r)^{2}}{r(\mathcal{L}_{X}c_s^{-2}+3F)}b^{2}\right],
\end{eqnarray}
\begin{eqnarray}\label{B}
  B=-\frac{HT}{u}\left[\frac{c}{\mathcal{L}_{X}c_s^{-2}+3F}-\frac{\mathcal{L}_{X}+3F+r}{r(\mathcal{L}_{X}c_s^{-2}+3F)}b
   \right.\nonumber\\ \left. +\frac{\mathcal{L}_X+9F}{(\mathcal{L}_{X}c_s^{-2}+3F)(1-3\frac{X}{M_p^2M^{2}})(\mathcal{L}_{X}+3F+r)^{2}}\epsilon\right],
\end{eqnarray}
\begin{eqnarray}\label{C}
C=\frac{3H^2 u}{T}\left[\frac{r}{(1-3\frac{X}{M_{p}^{2}M^{2}})(\mathcal{L}_{X}+3F+r)} \epsilon \right.\nonumber\\ \left. -(\mathcal{L}_{X}+3F+r)(1-c)b -\frac{r}{\mathcal{L}_{X}+3F+r}\beta\right],
\end{eqnarray}
\begin{equation}\label{D}
 D=\frac{Hu}{T}\left[6r-\frac{(\mathcal{L}_{X}c_s^{-2}+9F)r}{(1-3\frac{X}{M_{p}^{2}M^{2}})\left(\mathcal{L}_{X}
    +3F+r\right)^{2}}\epsilon\right],
\end{equation}
\begin{equation}\label{lambda1}
\lambda_{1}=-3H-3H\frac{r}{\mathcal{L}_{X}c_s^{-2}+3F}-H\epsilon\frac{(\mathcal{L}_X+9F)(\mathcal{L}_{X}c_s^{-2}+9F)}
{(\mathcal{L}_Xc_s^{-2}+3F)(1-3\frac{X}{M_{p}^{2}M^{2}}) (\mathcal{L}_{X}+3F+r)^2},
\end{equation}
\begin{equation}\label{lambda2}
 \lambda_{2}=-H(4-c)-H\frac{r}{(1-3\frac{X}{M_{p}^{2}M^{2}})(\mathcal{L}_{X}+3F+r)^{2}}\epsilon.
\end{equation}

The slow-roll solution will act as an attractor for the inflationary dynamical system only if the eigenvalues of matrix $E$ are either negative or positive but of sufficiently small magnitude $\mathcal{O}(\frac{\epsilon}{\mathcal{L}_X+3F+r})$ (i.e. with very slow growth). Additionally, ``forcing term'' $F$ must be sufficiently small, i.e., $|\frac{\dot u_0}{ H_0u_0}|$ , $|\frac{\dot s_0}{H_0s_0}| \ll 1$.
We first focus on the ``forcing term'' $F$. By taking the time derivatives of the slow-roll equations (\ref{SREOM}) and (\ref{SRentropy}), we obtain:
\begin{eqnarray}
\frac{\dot u_{0}}{Hu_{0}}&=& \frac{1}{\Delta}\left[\frac{(\mathcal{L}_X+9F)(4-c)-cr}{(\mathcal{L}_Xc_s^{-2}+3F)(1-3\frac{X}{M_{p}^{2}M^{2}})
(\mathcal{L}_{X}+3F+r)}\epsilon +\frac{4r}{(\mathcal{L}_Xc_s^{-2}+3F)(\mathcal{L}_X+3F+r)}\beta
\right.\nonumber\\ &-&\left.\frac{4-c}{\mathcal{L}_Xc_s^{-2}+3F}\eta
+\frac{\mathcal{L}_{X}+3F+r}{\mathcal{L}_{X}c_s^{-2}+3F}c(5-2c)b
-\frac{(\mathcal{L}_{X}+3F+r)^{2}}{r(\mathcal{L}_{X}c_s^{-2}+3F)}(13-4c)b^{2}\right],
\end{eqnarray}
\begin{eqnarray}
\frac{\dot s_{0}}{Hs_{0}}&=&
\frac{1}{\Delta} \left\{\frac{6(\mathcal{L}_X+9F)-3(\mathcal{L}_{X}c_s^{-2}+3F)+3r}
{(\mathcal{L}_{X}c_s^{-2}+3F)(1-\frac{3X}{M_{p}^{2}M^{2}})(\mathcal{L}_X+3F+r)}\epsilon+\frac{5r-3(\mathcal{L}_{X}c_{s}^{-2}+3F)}
{(\mathcal{L}_{X}c_s^{-2}+3F)(\mathcal{L}_{X}+3F+r)}\beta
\right.\nonumber\\ &-& \left.\frac{6}{\mathcal{L}_{X}c_s^{-2}+3F}\eta
+\frac{\mathcal{L}_{X}+3F+r}{\mathcal{L}_{X}c_s^{-2}+3F}[7c-1+3\frac{c-1}{r}(\mathcal{L}_Xc_s^{-2}+3F)]b\right\}.
\end{eqnarray}

The slow-roll regime also requires that $\frac{\dot T}{HT}=\frac{\dot s}{3Hs}-\frac{\mathcal{L}_{X}+3F+r}{r}b<<1$.
The quantity in the denominator of above equations $(1-\frac{3X}{M_{p}^{2}M^{2}})=(1-\frac{9FX}{\rho})$ is of order unity and the quantity $\Delta$ in the equations is defined as follows:
\begin{equation}
    \Delta\cong(4-c)\left(1+\frac{r}{\mathcal{L}_{X}c_s^{-2}+3F}\right)+\frac{2cr}{\mathcal{L}_{X}c_s^{-2}+3F}\sim \mathcal{O}(1).
\end{equation}
The Hubble parameter $\epsilon_{_{H}}$ should also vary slowly, i.e. $\epsilon_{_{H}}=-\frac{\dot H_0}{H_0^2}\simeq\frac{1}{\mathcal{L}_{X}+3F+r}\epsilon\ll1$.

The sufficient conditions for satisfying the slow-roll requirements are expressed as follows:
\begin{eqnarray}
    \epsilon\ll\mathcal{L}_{X}+3F+r, \quad \beta\ll\frac{(\mathcal{L}_{X}c_s^{-2}+3F)(\mathcal{L}_{X}+3F+r)}{r}, \nonumber\\ \eta\ll\mathcal{L}_{X}c_s^{-2}+3F,\quad\quad b\ll\frac{\mathcal{L}_{X}c_s^{-2}+3F}{\mathcal{L}_{X}+3F+r}.\quad\quad\quad
\end{eqnarray}

According to the above equations, the slow-roll conditions in our NMDC noncanonical warm inflationary model are significantly more relaxed compared to both canonical and noncanonical warm inflation in GR, as well as standard inflation in GR. Slow-roll inflation occurs naturally in this scenario, without the need to specify a concrete potential. Thus, the strict flatness requirement in cold inflation is effectively resolved. This relaxation is primarily due to the presence of two large overdamped terms: the gravitationally enhanced friction term and the thermal damping term in Eq. (\ref{EOM2}). Consequently, the selection of the potential becomes less restrictive, allowing for the incorporation of various new models into cosmological inflation. This is a direct outcome of the combined effects of thermal dissipation and the NMDC. Notably, these two damping terms interact competitively. Different to the traditional strength parameter $r$, we can define a more proper parameter $Q=\frac{\Gamma}{3H(\mathcal{L}_{X}+3F)}$ in this scenario to effectively describe the ratio of thermal dissipative damping to the other damping terms. When thermal dissipation dominates, i.e. $Q\gg1$, the NMDC warm inflation is in the strong dissipation regime and the behavior approximates canonical warm inflation in GR. By contrast, when $Q\ll1$, thermal effects remain relatively weak, but still distinct from cold noncanonical inflation, a point that will be further elaborated later in this paper. The slow-roll condition for $b$ implies that the thermal correction to the inflationary potential should remain small, as is the case for warm inflation in GR \cite{Ian2008,Campo2010,Zhang2021}. Thus, the total energy density can take a nearly separable form $\rho\simeq\rho(\phi,X)+9FX+\rho_r$.

Matrix $E$ is used to derive the remaining slow-roll condition. From the slow-roll conditions established earlier, we obtain:
\begin{eqnarray}\label{charEq}
\det(\lambda I-E) &= &\left|
\begin{array}{ccc}
\lambda  & -1 & 0 \\
-A & \lambda-\lambda_1  & -B \\
-C & -D & \lambda-\lambda_2
\end{array}
\right|   \nonumber \\
&=&\lambda (\lambda-\lambda_1 )(\lambda-\lambda_2 )-BD\lambda-A(\lambda-\lambda_2)-BC   \nonumber
\\
&=&0,
\end{eqnarray}
where a very small eigenvalue $\lambda \simeq \frac{BC-A\lambda _2}{\lambda _1\lambda _2-BD-A}\ll\lambda_1,\lambda_2$ exists. The other two eigenvalues satisfy $\lambda ^2-(\lambda _1+\lambda _2)\lambda +\lambda _1\lambda _2-BD=0$. These two eigenvalues are both negative when $\lambda _1+\lambda _2<0$ and $\lambda_1\lambda _2-BD>0$. Finally the relation can be get:
\begin{equation}
|c|<4.  \label{c}
\end{equation}

The radiation energy density remains subdominant during the slow-roll inflationary epoch, i.e. $\frac{\rho_{r}}{V}=\frac{r\epsilon}{2(\mathcal{L}_{X}+3F+r)^2}\ll 1$, consistent with the prior assumption that inflation is potential-dominated.

\section{\label{sec:level4} Cosmological perturbations}

The development of cosmological perturbation theory within the NMDC warm noncanonical inflationary framework is now considered. Unlike in cold inflation, where density fluctuations arise primarily from quantum fluctuations, in warm inflation these fluctuations originate mainly from thermal fluctuations. Both entropy and curvature
perturbations may exist in warm inflationary scenarios. However, since radiation energy is subdominant
and its fluctuations contribute only to entropy perturbations, which decay on large scales \cite{Ian2008,Cai2011,Lisa2004}, the focus is placed solely on curvature perturbations, which persist at large scales.

The full inflaton field is expanded as $ \Phi(\mathbf{x},t)=\phi(t)+\delta\phi(\mathbf{x},t)$, accounting for small perturbations, where $\delta\phi(\mathbf{x},t)$ represents the linear response due to thermal stochastic noise $\xi$ within the thermal system. In the high-temperature limit $T\rightarrow\infty$, the noise source becomes Markovian: $\langle\xi(\mathbf{k},t)\xi(-\mathbf{k'},t')\rangle=2\Gamma Ta^{-3}(2\pi)^3\delta^3(\mathbf{k}-\mathbf{k'})\delta(t-t')$ \cite{Lisa2004,Gleiser1994}. By introducing a thermal white noise term and deriving from the full inflaton field equation, Eq. (\ref{EOM0}), the second-order Langevin equation for the filed is obtained:

\begin{eqnarray}\label{pEOM1}
    (\mathcal{L}_{X}c_{s}^{-2}+3F)(\ddot{\phi}(t)+\delta\ddot{\phi}(\mathbf{x},t))+
    \left[3H(\mathcal{L}_{X}+3F)+\Gamma\right]
    (\dot{\phi}(t)+\delta\dot{\phi}(\mathbf{x},t))+V_{\phi}\nonumber\\+V_{\phi\phi}\delta\phi(\mathbf{x},t)-
    \mathcal{L}_{X}\frac{\nabla^{2}}{a^{2}}\delta\phi(\mathbf{x},t)+3Fw\frac{\nabla^{2}}{a^{2}}\delta\phi(\mathbf{x},t)
    =\xi(\mathbf{x},t).\quad\quad\quad\quad
\end{eqnarray}
Applying the Fourier transform, the evolution equation for the fluctuation is derived:
\begin{equation}\label{pEOM2}
    (\mathcal{L}_{X}c_{s}^{-2}+3F)\delta\ddot{\phi}_{\mathbf{k}}+[3H(\mathcal{L}_{X}+3F)+
    \Gamma]\delta\dot{\phi}_{\mathbf{k}}
    +\left(\mathcal{L}_{X}\frac{k^{2}}{a^{2}}-3Fw\frac{k^{2}}{a^{2}}+m^{2}\right)\delta\phi_{\mathbf{k}}=\xi_{\mathbf{k}},
\end{equation}
where the term $m^2=V_{\phi\phi}$ is the effective squared inflaton mass. Although second-order Langevin equations are difficult to solve, the power spectrum can be calculated at the point of horizon crossing. Horizon crossing occurs well within the slow-roll inflationary regime \cite{LiddleLyth}, which is typically overdamped, especially in this case as proved in Section \ref{sec:level3}, allowing the inertia term to be neglected. In the slow-roll regime, the state parameter simplifies to $w=\frac{p}{\rho}\simeq-1$. The Langevin equation (\ref{pEOM2}) is thus reduced to a first-order equation, as shown in \cite{Berera2000,TalyorBerera}:
\begin{equation}\label{pEOM3}
[3H(\mathcal{L}_{X}+3F)+\Gamma]\delta\dot{\phi_{\mathbf{k}}}+\left(\mathcal{L}_{X}\frac{k^2}{a^2}+3F\frac{k^{2}}{a^{2}}
+m^2\right)\delta\phi_{\mathbf{k}}=\xi_{\mathbf{k}}.
 \end{equation}

In the deep slow-roll regime, an approximate analytic solution is given by:
\begin{eqnarray}\label{phik}
    \delta\phi_{\mathbf{k}}(t)&\simeq&\frac{1}{3H(\mathcal{L}_{X}+3F)+\Gamma}\exp\left[-\frac{t-t_{0}}{\tau(\phi_0)}\right]
    \int_{t_{0}}^{t}\exp\left[\frac{t^{\prime}-t_{0}}{\tau(\phi_0)}\right]\xi(\mathbf{k},t^{\prime})dt' \nonumber\\
    &+&\delta\phi(\mathbf{k},t_{0})
    \exp\left[-\frac{t-t_{0}}{\tau(\phi_0)}\right],
\end{eqnarray}

where $\tau(\phi)=\frac{3H(\mathcal{L}_{X}+3F)+\Gamma}{(\mathcal{L}_{X}+3F)\frac{k^{2}}{a^{2}}+m^2}$ describes the efficiency of the thermalizing process. The relation between the physical wavenumber $k_p$ and comoving wavenumber $k$ is $k_p=k/a$. According to Eq. (\ref{phik}), the larger $k_p^2$, the faster the relaxation rate. A mode can thermalize if $k_p^2$ is sufficiently large for the mode to relax within a Hubble time. As the universe expands, once the physical wavenumber of the field mode $\delta\phi(\mathbf{x},t)$ falls below $k_F$, it no longer experiences the effects of thermal noise $\xi(\mathbf{k},t)$ during a Hubble time \cite{Berera2000}. Based on this criterion, the freeze-out physical momentum $k_F$ is defined as $\tau(\phi)=\frac{1}{H}$, i.e. $\frac{(\mathcal{L}_{X}+3F)k_{F}^{2}+m^2}{3H^{2}(\mathcal{L}_{X}+3F+r)}=1$.
In the slow-roll regime, the mass term is negligible compared to other terms, leading to:
\begin{equation}
    k_F=H\sqrt{\frac{3(\mathcal{L}_X+3F+r)}{\mathcal{L}_X+3F}}=H\sqrt{3(1+Q)}.
\end{equation}
In the strong dissipative regime of warm inflation, where $Q\gg1$, $k_F$ exceeds $H$, indicating that freeze-out occurs significantly earlier than horizon crossing. In the weak dissipative regime, where $Q\ll1$, $k_F$ remains slightly larger than $H$.

Using the field perturbation relation $\delta\phi^2=\frac{k_FT}{2\pi^2}$ in warm inflation \cite{Berera2000,TalyorBerera}, and working in the spatially flat gauge with $\mathcal{R}=\frac{H}{\dot\phi}\delta\phi_{\mathbf{k}}$, the scalar power spectrum for the NMDC warm noncanonical inflationary model is derived as follows:
\begin{eqnarray}\label{PR}
P_{R}&=&(\frac{H}{\dot{\phi}})^{2} \delta\phi^{2}
=\frac{H^{3}T}{2\pi^{2}u^{2}}\sqrt{\frac{3(\mathcal{L}_{X}+3F+r)}{\mathcal{L}_{X}+3F}}\nonumber\\
&=&\frac{9H^{5}T\left(\mathcal{L}_{X}+3F+r\right)^{\frac52}}{2\pi^{2}V_{\phi}^{2}}\sqrt{\frac{3}{\mathcal{L}_X+3F}}.
\end{eqnarray}

CMB observations often provide a reliable normalization for the scalar power spectrum $P_R\approx 10^{-9}$ on large scales. The term $(\mathcal{L}_X+3F+r)^{5/2}$ in the numerator indicates that the energy scale at horizon crossing is significantly suppressed by both the nonminimal coupling effect and thermal dissipation. This result supports the assumption that inflation can be effectively described by field theory. The spectral index $n_s$ for the power spectrum is given by:

\begin{eqnarray}
n_s-1&=&\frac{d\ln P_{R}}{d\ln k}\simeq\frac{\dot P_{R}}{HP_{R}}\nonumber\\
&=&\alpha_{1}\frac{\epsilon}{\mathcal{L}_{X}+3F+r}+\alpha_{2}\frac{\eta}{\mathcal{L}_{X}c_s^{-2}+3F}
+\alpha_{3}\frac{r\beta}{(\mathcal{L}_{X}c_s^{-2}+3F)(\mathcal{L}_{X}+3F+r)}\nonumber\\ &+&\alpha_{4}\frac{\mathcal{L}_{X}+3F+r}
{\mathcal{L}_{X}c_s^{-2}+3F}b.
\end{eqnarray}
In the calculations, the parameters $\alpha_1$, $\alpha_2$, $\alpha_3$ and $\alpha_4$ are defined as follows:
\begin{eqnarray}
\alpha_{1}&=&\left\{-3-\frac{3F}{\mathcal{L}_{X}+3F+r}+\frac{3F}
{\mathcal{L}_{X}+3F}-\frac{r}{2(\mathcal{L}_{X}+3F+r)}\right\}\nonumber\\
&+&\frac{1}{\Delta}\left\{ \left[\frac{1}{3}-\frac{rc}{6(\mathcal{L}_{X}+3F+r)}\right]\frac{6(\mathcal{L}_X+9F)-3(\mathcal{L}_Xc_s^{-2}+3F)+3r}
{\left(1-3\frac{X}{M_{p}^{2}M^{2}}\right)(\mathcal{L}_Xc_s^{-2}+3F)}
\right. \\ &+& \left.\left[\left(\frac{\mathcal{L}_{X}}{2(\mathcal{L}_{X}+3F+r)}-\frac{\mathcal{L}_{X}}{2(\mathcal{L}_{X}+3F)}\right)
(c_{s}^{-2}-1)-2\right]\frac{(\mathcal{L}_X+9F)(4-c)-cr}{(\mathcal{L}_{X}c_s^{-2}+3F)(1-\frac{3X}{M_{p}^{2}M^{2}})} \right\},\nonumber
\end{eqnarray}
\begin{eqnarray}
    \alpha_{2}&=&\frac{1}{\Delta}\left\{\frac{rc}{\mathcal{L}_X+3F+r}-2-(4-c)\right.\nonumber\\ &\times & \left.
    \left[\left(\frac{\mathcal{L}_{X}}{2(\mathcal{L}_{X}+3F+r)}-\frac{\mathcal{L}_{X}}{2(\mathcal{L}_{X}+3F)}\right)
(c_{s}^{-2}-1)-2\right]\right\},
\end{eqnarray}
\begin{eqnarray}
\alpha_{3}&=&\frac{1}{\Delta}\left\{\left[\frac{1}{3}-\frac{rc}{6(\mathcal{L}_{X}+3F+r)}\right]
    \left[5-\frac{3}{r}(\mathcal{L}_{X}c_s^{-2}+3F)\right]\right.\nonumber\\ &+&\left.\left[\frac{2\mathcal{L}_{X}}{\mathcal{L}_{X}+3F+r}
    -\frac{2\mathcal{L}_{X}}{\mathcal{L}_{X}+3F}\right]\left(c_s^{-2}-1\right)-8\right\}
-\frac{\mathcal{L}_{X}c_s^{-2}+3F}{2(\mathcal{L}_{X}+3F+r)},
\end{eqnarray}
\begin{eqnarray}
\alpha_{4}&=&\frac{1}{\Delta}\left\{\left[7c-1+3(\mathcal{L}_{X}c_s^{-2}+3F)\frac{c-1}{r}\right] \left[\frac{1}{3}-\frac{rc}{6(\mathcal{L}_{X}+3F+r)}\right]
\right.\nonumber\\ &+&\left.\left[\left(\frac{\mathcal{L}_{X}}{2(\mathcal{L}_{X}+3F+r)}-\frac{\mathcal{L}_{X}}
{2(\mathcal{L}_{X}+3F)}\right)(c_{s}^{-2}-1)-2\right]c(5-2c) \right\}
\\&+&\frac{\mathcal{L}_{X}c_{s}^{-2}+3F}{\mathcal{L}_{X}+3F+r}\left(\frac{c}{6}-\frac{\mathcal{L}_{X}+3F+r}{r}\right).\nonumber
\end{eqnarray}

These parameters are all of order unity, leading to the conclusion that $(n_s-1)$ is a first-order small quantity in slow-roll regime, specifically of the order $\mathcal{O}\left(\frac{\epsilon}{\mathcal{L}_X+3F+r}\right)\ll1$. Consequently, a nearly scale-invariant power spectrum is obtained, in agreement with observations. The running of the spectral index $\alpha_s=\frac{dn_s}{d\ln k}$ is of the order $\left(\frac{\epsilon}{\mathcal{L}_X+3F+r}\right)^2\ll (n_s-1)$, which is also consistent with observational data. This NMDC noncanonical warm inflationary scenario thus can be theoretically fit observations. In future work, specific models within this scenario will be numerically analyzed, and physical model quantities will be determined by detailed comparison with the latest {\it Planck} observations.

Tensor perturbations do not couple to the thermal background and are generated solely by quantum fluctuations, as in standard inflation \cite{TalyorBerera}. The nonminimal derivative coupling $G^{\mu\nu} \partial_{\mu}\phi\partial_{\nu}\phi$ introduces only a small correction to the power spectrum of gravitational waves \cite{DalianisJCAP2020,Karydas2102.08450}:
\begin{equation}\label{tensorperturbation}
    P_T\simeq\frac{2}{M_p^2}\left(\frac{H}{2\pi}\right)^2.
\end{equation}
To the first order of approximation, the spectral index of tensor perturbations is expressed as  $n_T\simeq-2\epsilon_{_H}=\frac{2\epsilon}{\mathcal{L}_X+3F+r}$. The tensor-to-scalar ratio $R$ is expressed as follows:
\begin{equation}
R=\frac{P_{T}}{P_{R}}=\frac{H}{T}\frac{2\epsilon(\mathcal{L}_X+3F)^{\frac12}}{\sqrt{3}(\mathcal{L}_{X}+3F+r)^{\frac52}}.
\end{equation}
Given that the scalar power spectrum is fixed by observations, tensor perturbations can be significantly weakened due to the combined effects of nonminimal derivative coupling and thermal dissipation, particularly when both effects are strong. By considering the slow-roll condition $\epsilon<
\mathcal{L}_X+3F+r$, the upper bound for the tensor-to-scalar ratio is determined as
\begin{equation}
R\leq\frac{H}{T}\frac{2(\mathcal{L}_{X}+3F)^{\frac{1}{2}}}{\sqrt{3}(\mathcal{L}_{X}+3F+r)^{\frac{3}{2}}}.
\end{equation}
The contribution to the best-fit tensor-to-scalar ratio corresponds to a value near the observational limit $R=(3-6)\times 10^{-3}$ \cite{BICEP2}, with the upper bound of $R$ being sufficiently small to align with current constraints. These findings suggest that a small $R$ is just right favored in the NMDC warm noncanonical inflationary scenario.

The insignificant non-Gaussianity indicated by {\it Planck} observations \cite{PLANCK2} also favors a relatively large sound speed and moderate thermal dissipation in the noncanonical warm inflation model \cite{Cai2011,Franche2010}. A more detailed analysis of non-Gaussian features and the observational constraints on the parameters of the NMDC noncanonical warm inflation model will be the focus of future work.

The amount of expansion is $\Delta N\simeq 4.6$ while the scales corresponding to $1\leq l \leq100$ are leaving the horizon, and the corresponding field variation is as follows:
\begin{equation}
\frac{\Delta\phi}{M_{p}}=\frac{\dot{\phi}\Delta N}{M_{p}H}\lesssim\frac{4.6\sqrt{2}}{(\mathcal{L}_{X}+3F+r)^{\frac12}}\sim (\mathcal{L}_{X}+3F+r)^{-\frac12}.
\end{equation}
The standard Lyth bound in general relativity (GR) \cite{LythPRL1997} is modified in this context, indicating that the field excursion is reduced to well below the Planck scale ($\Delta\phi\ll M_p$) in the NMDC warm noncanonical inflationary model, contrary to standard inflation ($\frac{\Delta\phi}{M_p}=0.5(\frac{R}{0.1})^{1/2}$, where a detectable $R$ requires a Planckian or even super-Planckian inflaton excursion \cite{LiddleLyth}). Thus, the issue of an excessively large inflaton amplitude is naturally resolved in this NMDC warm noncanonical inflationary scenario, regardless of the potential type. The consistency relation becomes:
\begin{equation}
R=-\frac HT\frac{(\mathcal{L}_{X}+3F)^{\frac12}n_{T}}{\sqrt{3}(\mathcal{L}_X+3F+r)^{\frac32}}.
\end{equation}
This is no longer a fixed relation, as in minimal cold inflation ($R=-6.2n_T$ \cite{LiddleLyth}).

The radiation energy density and the universal temperature follow the Stefan-Boltzmann relationship $\rho_r=\pi^2g_{\ast}T^4/30$. By combining the slow-roll equations and the power spectrum, the following result is obtained:
\begin{equation}\label{TH}
\frac{T}{H}
    =(\frac{45r}{4\pi^{4}g_{*}P_{R}})^{\frac{1}{3}}\quad\left[\frac{3(\mathcal{L}_{X}+3F+r)}{\mathcal{L}_{X}+3F}\right]
    ^{\frac{1}{6}}.
\end{equation}
The ratio $T/H$ is slightly smaller than in canonical warm inflation or noncanonical warm inflation in the GR limit \cite{Ian2008,Zhang2014,Zhang2021}. However, an increase in thermal dissipation strength can enhance the $T/H$ ratio. When strong thermal dissipation dominates gravitational damping, the universal bath temperature rises. Conversely, weak dissipation leads to a reduction in temperature since the thermal effect competes with the NMDC effect. The condition for warm inflation to occur, $T>H$, can be easily satisfied when $r>g_{\ast}P_R$, as indicated by Eq. (\ref{TH}). Given that $g_{\ast}$ is of the order of $\mathcal{O}(10^2)$ during inflation and $P_R$ is of the order of $\mathcal{O}(10^{-9})$, even very weak thermal dissipation can initiate warm inflation. Thus, warm inflation may offer a more realistic description of the early accelerating universe. Moreover, even in the weak dissipative regime $Q<1$, thermal fluctuations continue to dominate over quantum fluctuations, as a direct consequence of Eqs. (\ref{PR}) and (\ref{TH}).

\section{\label{sec:level5}Conclusions and discussions}
This paper proposed a model of nonminimal derivative coupling (NMDC) warm noncanonical inflation, expanding the range of inflationary scenarios. By analyzing the total action of the NMDC warm inflationary system, the evolution equations for both the inflaton and radiation were derived, along with other key equations for this new framework. A significant feature of the model was the greatly enhanced damping term, resulting from both NMDC and thermal dissipation effects. This enhancement made it much easier to satisfy the slow roll approximations, which represented one of the most attractive aspects of this approach.

The stability analysis revealed a set of relaxed slow roll conditions, characterized by warm inflationary slow roll parameters. By analytically examining the perturbation evolution equations, a new form of the scalar power spectrum was derived that remained nearly scale-invariant and qualitatively fit observational data. Moreover, the energy scale at horizon crossing was significantly reduced due to the combined effects of gravitationally enhanced friction and thermal dissipation. The amplitude of gravitational waves in the model was weaker than that in warm inflation under general relativity and even more so compared to cold inflationary models.

Warm noncanonical inflation with a nonminimal derivative coupling term offered a promising solution to the problem of excessively large inflaton amplitudes, ensuring that the inflaton field excursion remained well below the Planck scale. Additionally, analysis of the relation of temperature revealed that very weak thermal dissipation could result in warm inflation.

Future research should focus on exploring specific models within this new framework to provide more precise comparisons with observational data. Furthermore, the non-Gaussian features in this scenario represents an area that must be comprehensively investigated.

%----------------------------------------------------------------------------------------------------
\acknowledgments This work was supported by projects ZR2021MA037, ZR2022JQ04 and ZR2021QA037 supported by Shandong Provincial Natural Science Foundation, National Natural Science Foundation of China (Grant No. 11605100 and No. 12205158), the Natural Science Foundation of Henan under Grant No. 232300421351 and the Talent Introduction Fund (Grant No. 2020BS035) at Henan University of Technology.
%---------------------------------------------------------------------------------------------------------

\end{document}